# 2D and 3D Computer Graphics Algorithms under MorphoSys

Issam Damaj, Suhaib Majzoub and Hassan Diab

issamwd@ieee.org; s_majzoub@hotmail.com; diab@aub.edu.lb
Department of Electrical and Computer Engineering
Faculty of Engineering and Architecture
American University of Beirut
P.O. Box 110236
Beirut, Lebanon

**Abstract.** This paper presents new mappings of 2D and 3D geometrical transformation on the MorphoSys (M1) reconfigurable computing (RC) prototype [2]. This improves the system performance as a graphics accelerator [1-5]. Three algorithms are mapped including two for calculating 2D transformations, and one for 3D transformations. The results presented indicate an improved performance. The speedup achieved is explained as well as the advantages in the mapping of the application. The transformations on an 8x8 RC array were run, and numerical examples were simulated to validate our results, using the MorphoSys mULATE program, which simulates MorphoSys operations. Comparisons with other systems are presented, namely, with Intel processing systems and Celoxica RC-1000 FPGA.

## 1 Introduction

Reconfigurable computing (RC) is becoming more popular and increasing research efforts are being invested in it. It employs reconfigurable hardware and programmable processors. The application is mapped such that the workload is divided between the general-purpose processor and the reconfigurable device. The use of RC opens the way for an increased speed over general-purpose processors and a wider functionality than application specific integrated circuits (ASICs). It is a good solution for applications requiring a wide range of functionality and speed at the same time [1].

## 2 MorphoSys Design

One of the emerging RC systems includes the MorphoSys designed and implemented at the University of California, Irvine. It is composed of: 1) an array of reconfigurable cells called the RC array, 2) its configuration data memory called context memory, 3) a control processor (TinyRISC), 4) a data buffer called the frame buffer, and 5) a DMA controller [2].

A program runs on MorphoSys in the following manner: General-purpose operations are handled by the TinyRISC processor, while operations that have a certain degree of parallelism, regularity, or intensive computations are mapped to the reconfigurable array (RC-Array).

## 3 Geometrical Transformations

### 3.1 First 2D Algorithm Mapping

The main usage of the MorphoSys is, as any parallel processor, to perform fast computations of algorithms that need a certain computational power requirement. Computer graphics algorithms represent one of these families. A basic part in computer graphics operations is geometrical transformations, which require fast computations of matrix operations, namely, matrix multiplication which is the core part of any geometrical transformation. The emphasis in this paper is the mapping of matrix multiplication on the MorphoSys for the use with computer graphics 2D transformations, using the supported internal configuration of the RCs.

This Algorithm could be mapped onto the M1 RC-Array as follows: The contents of the matrix A are passed row by row through the context words. The contents of matrix B are broadcasted row by row to the columns of the RC-Array. The multiplication stage (row x column) is done by using the CMUL ALU operation where the output of the reconfigurable cell is: Out (t) = A x B. Then, the results output from the ALU of each RC need to be accumulated in a row-wise manner so we can get the first row of the output matrix C from the last column of the RC-Array. This is done by using the ALU-operation CMULOADD where Out (t + 1) = Out (t) + Out [From Left Cell]. Indeed, the contents of column 7 of the RC-Array are stored back to the frame memory and then to the main memory. The same steps are repeated with the same context word but with different constant field containing the data from matrix A until obtaining the resultant matrix C.

### 3.2 Second 2D Algorithm Mapping

Using the same above algorithm, in this section, we introduce a new mapping, taking the advantages of the MorphoSys reconfigurable array topologies. The new mapping uses the upper left quadrant along with the bottom right quadrant. Where, the matrices are considered to be of size 4x4.

### 3.3 3D Geometrical Transformations

The basic purpose of composing transformations is to gain efficiency by applying a single composed transformation to a point, rather than applying a series of transformations, one after the other. Our proposed mapping assumes column broadcast mode where all the cells in the same column perform the same function. The desired functions of the interconnection are: Out(t+1) = A×C,  Out(t+1) = A×C + Out(t), and Out(t+1) = A×C – Out(t).

## 4   Performance Analysis

The performance is based on the execution speed of the algorithms. The MorphoSys system is considered to be operational at a frequency of 100 MHz. Comparisons among the suggested systems are given in Table 1 recalling some previous findings from [4-5], showing the speedup factor of the M1 over the other suggested systems. The speedup factor is calculated as the ratio in number of cycles between the M1 and the other suggested systems. Comparisons with the RC-1000 Celoxica FPGA are shown in Table 2.

**Table 1.** Comparisons with other systems.

| Algorithm | System | N# of Cycles | Speedup |
|---|---|---|---|
| **Vector-Vector Operations (Translation) "64-Elements".** | M1 | 96 | |
| | 80486 | 769 | 8.01 |
| **Vector-Scaling Operations (Scaling) "64-Elements".** | M1 | 55 | |
| | 80486 | 578 | 10.5 |
| **Combined Translations "64-Elements".** | M1 | 150 | |
| | 80486 | 898 | 5.99 |
| **Combined Translation and Scaling "64-Elements".** | M1 | 96 | |
| | 80486 | 3395 | 62.62 |
| **Combined Scaling "64-Elements".** | M1 | 66 | |
| | Pentium | 2053 | 31.1 |
| | 80486 | 6147 | 93 |
| **General Composite Algorithm Using Matrix Algorithm "64-Elements". Algorithm 1.** | M1 | 256 | |
| | Pentium | 10151 | 39.65 |
| | 80486 | 27038 | 105.61 |
| **General Composite Algorithm Using Matrix Algorithm "16 Elements". Algorithm 2.** | M1 | 70 | |
| | Pentium | 1328 | 18.97 |
| | 80486 | 3354 | 47.91 |
| **General Composite Algorithm Using Matrix Algorithm "64-Elements". Algorithm 3.** | M1 | 45 | |
| | Pentium | 2551 | 56.67 |
| | 80486 | 6773 | 150.5 |

The link to the formal publication is via
https://doi.org/10.1007/3-540-46117-5_111Table 2. Comparisons with RC-1000 FPGA.

| Algorithm | System | N# of Cycles | Speedup of the RC-1000 over the M1 RC-System. |
|---|---|---|---|
| **General Composite Algorithm Using Matrix Algorithm "64-Elements". Algorithm 1.** | M1 | 256 | |
| | RC-1000 | 15 | 17 |
| **General Composite Algorithm Using Matrix Algorithm "16 Elements". Algorithm 2.** | M1 | 70 | |
| | RC-1000 | 12 | 5.8 |
| **General Composite Algorithm Using Matrix Algorithm "64-Elements". Algorithm 3.** | M1 | 45 | |
| | RC-1000 | 12 | 3.7 |

## 5   Conclusion

New mapping techniques for some linear algebraic functions are recalled [4-5]. New mappings for 2D and 3D geometrical transformations are introduced and justified dealing with transformations operations and their performance analysis under MorphoSys. The speed of this mapping is calculated. Results compared with other processing systems. The superiority of the presence of a reconfigurable coprocessor is apparent from the calculated speedups. Effort could be invested in trying to map other algorithms that make use of the mapped ones for more advanced algorithms for computer graphics.

## References

1. Abdennour E., Diab, H., Kurdahi, F.: FIR Filter Mapping and Performance Analysis on MorphoSys. *Proceedings of the Seventh IEEE International Conference on Electronics, Circuits and Systems.* Kaslik, Lebanon. (17-20 December 2000) 99-102.
2. Maestre R., Kurdahi, F., Bagherzadeh, N., Singh, H., Hermida, R., Fernandez, N.: Kernel Scheduling in Reconfigurable Computing. *Proceedings of Design and Test in Europe* (DATE'99). Munich, Germany, March 1999.
3. Bagherzadeh N., Kurdahi, F., Singh, H., Lu, G., Lee, M., Filho, E.: MorphoSys: A Reconfigurable Architecture for Multimedia Applications. *Proceedings of the XI Brazilian Symposium on Integrated Circuit Design.* Rio De Janeiro, October 1998.
4. Damaj I., Diab, H.: Performance Analysis of Extended Vector-Scalar Operations Using Reconfigurable Computing. *Proceedings of the ACS/IEEE International Conference on Computer Systems and Applications*. Beirut, Lebanon. (25-29 June 2001) 227-232.
5. Damaj I., Diab, H.: Graphics Acceleration Using Reconfigurable Computing. *Proceedings of the 13th International Conference on Microelectronics* Rabat, Morocco (29-31 October 2001).